\documentclass[10pt,conference]{IEEEtran}

\usepackage[utf8]{inputenc} 
\usepackage[T1]{fontenc}
\usepackage{url}
\usepackage{ifthen}
\usepackage[cmex10]{amsmath} 

\usepackage{eurosym}
\usepackage{amssymb}
\usepackage{amsfonts}
\usepackage{float}
\usepackage{graphics}
\usepackage[tight,footnotesize]{subfigure}
\usepackage{rotating}
\usepackage{algorithm}
\usepackage{enumerate}
\usepackage{cite}

\usepackage{xcolor}
\usepackage{color,soul}
\usepackage[colorinlistoftodos]{todonotes}
\usepackage{lipsum}
\usepackage{cases}
\usepackage{comment}
\usepackage{algpseudocode}
\usepackage{resizegather}

\usepackage{multirow}
\usepackage{threeparttable}
\usepackage{arydshln}

\usepackage{mathtools}

\DeclareMathSizes{10}{9}{7}{5}
\newtheorem{theo}{Theorem}

\newtheorem{theorem}{Theorem}

\newtheorem{definition}[theorem]{Definition}

\newtheorem{lemma}{Lemma}

\newtheorem{remark}{Remark}

\newlength{\myeqskip}  

\expandafter\def\expandafter\normalsize\expandafter{%
    \normalsize%
    \setlength\abovedisplayskip{\myeqskip}%
    \setlength\belowdisplayskip{\myeqskip}%
    \setlength\abovedisplayshortskip{\myeqskip}%
    \setlength\belowdisplayshortskip{\myeqskip}%
}

\allowdisplaybreaks[4]

\begin{document}
\title{Joint Power and Bit Allocation for Precoded Massive MIMO Channels} 



    \author{\IEEEauthorblockN{Shuiyin  Liu}
\IEEEauthorblockA{{Cyber Security Research and Innovation Centre} \\
{Holmes Institute}\\ {Melbourne, VIC 3000, Australia} \\
Email: SLiu@Holmes.edu.au}
\and
\IEEEauthorblockN{Amin Sakzad}
\IEEEauthorblockA{
{Faculty of Information Technology}\\
{Monash University}\\ {Melbourne, VIC 3800, Australia} \\
Email: Amin.Sakzad@monash.edu}
}

\maketitle


\begin{abstract}
This work addresses the joint optimization of power and bit allocation in precoded large-scale \( n \times n \) MIMO systems with discrete input alphabets, specifically \( \{M_i\}_{i=1}^{n} \)-QAM constellations. We propose an adaptive QAM scheme that maintains a fixed gap to the Gaussian-input capacity for a given \( n \). A key finding is that, under the proposed scheme, the mercury/waterfilling (MWF) solution reduces analytically to the classical water-filling (WF) policy. Furthermore, the adaptive QAM configuration can be precomputed under the large-system assumption, enabling the replacement of full SVD with truncated SVD and yielding substantial computational savings. To support practical deployment, we develop a bit-allocation algorithm that meets a target transmission data rate while minimizing the overall decoding error rate and preserving computational complexity at \( O(n \log n) \). Simulation results confirm that the proposed truncated SVD precoding, paired with the joint power–bit allocation, achieves superior decoding performance relative to conventional approaches, while operating at significantly lower complexity.

\end{abstract}

\section{Introduction}
Massive Multiple-Input Multiple-Output (MIMO) has emerged as a key enabler for satisfying the performance requirements of 5G and future 6G wireless networks. With channel state information available at both the transmitter (CSIT) and the receiver (CSIR), massive MIMO systems can exploit spatial multiplexing and advanced beamforming/precoding schemes to support the simultaneous transmission and reception of multiple data streams across multiple user terminals \cite{MMIMO2019}. However, the computational burden associated with precoding and detection increases with the increasing number of antennas. Consequently, the design of low-complexity precoding and decoding algorithms is critical for the practical deployment of massive MIMO systems.

In multi-user (MU) MIMO, low-complexity linear precoding (e.g., ZF \cite{ZFPrecoding2006}) incurs significant performance loss, while non-linear schemes like vector perturbation \cite{Hochwald05} and LR-aided precoding \cite{Windpassinger04} achieve better performance. The idea of precoding schemes \cite{Hochwald05} \cite{Windpassinger04} is to transform the MIMO system into parallel subchannels that have the same signal-to-noise ratio (SNR). This design concept can also be interpreted as placing the largest signal components along the smallest singular values of the inverse channel matrix, and the smallest signal components along the largest singular values of the inverse channel \cite{Hochwald05}. However, the matrix representing a large inverse channel is ill-conditioned \cite{Tao2010}, meaning that the smallest singular value tends to zero as the number of antennas increases. In \cite{Shuiyin12}, it shows that the performance of LR aided precoding degrades exponentially as the number of antennas increases.

Block diagonalization \cite{SVD_BD_2004} enables the extension of single-user (SU) MIMO precoding methods, such as SVD-based precoding \cite{Tse05}, to the MU-MIMO context. In terms of power allocation, LR-aided precoding \cite{Hochwald05}\cite{Windpassinger04} adopts a strategy distinct from the classical waterfilling (WF) approach used in SVD-based precoding, which is capacity-optimal under Gaussian input assumptions. WF allocates less power to subchannels with smaller singular values, whereas LR-aided precoding does not exploit this property, partly explaining its suboptimal performance. Although Gaussian inputs maximize mutual information theoretically, they are infeasible in practical systems where inputs are drawn from discrete constellations (e.g., \( M_i \)-QAM constellation for the $i^{\text{th}}$ subchannel). In such cases, the optimal power allocation is governed by the \emph{mercury/waterfilling} (MWF) principle \cite{MWF2006}, which maximizes mutual information across the \( n \) subchannels. Unlike WF, MWF lacks a closed-form solution, though accurate approximations have been proposed (e.g., \cite{MWF_App2021}). An alternative power allocation scheme is the \emph{error/waterfilling} (EWF) strategy \cite{EWF2003}, originally developed for OFDM systems to minimize total decoding error probability. Both MWF and EWF require a priori specification of the input constellation sizes \( \{M_i\}_{i=1}^{n} \) to compute optimal power distribution. While various adaptive QAM schemes (A-QAM) have been proposed in the literature \cite{MWF_App2021}\cite{Palomar2005}, a comprehensive understanding of the inter-dependencies among QAM size selection, bit-allocation, and power allocation, and their collective effect on overall decoding performance, remains an open research challenge.

The primary contribution of this work lies in the joint design of power and bit allocation for precoded massive MIMO systems with discrete input constellations. First, we demonstrate that advanced power allocation strategies—whether aimed at maximizing mutual information (MWF) or minimizing total error rate (EWF)—consistently deactivate \( k_{\text{opt}} > 0 \) of the weakest subchannels under an average power constraint, assuming a sufficiently large \( n \). Under the large-system assumption \cite[Theorem 2.37]{Verdu04}, \( k_{\text{opt}} \) can be approximated by a constant, enabling the replacement of full SVD with \emph{truncated SVD} and yielding substantial computational savings. Second, we introduce an adaptive QAM scheme that maintains a constant gap to the Gaussian-input capacity for a given $n$. Notably, we show that under the proposed adaptive QAM approach, the MWF strategy analytically reduces to the classical WF solution, which is of independent theoretical interest. To support practical implementation, we further develop a bit-allocation algorithm tailored to a specified data transmission rate,  with the dual objective of minimizing the overall decoding error rate and maintaining computational complexity at \( O(n\log n) \). Simulation results confirm that combining the truncated SVD precoding with the proposed joint power and bit allocation approach enhances decoding performance while lowering precoding complexity relative to conventional methods.

\emph{Organization:} Section II introduces the notation and describes the channel model. Section III analyzes the number of deactivated subchannels under WF, MWF, and EWF. Section IV outlines the proposed adaptive QAM scheme along with the bit-allocation algorithm. Section V concludes the key findings.

\section{System Model}
In this section we provide some definitions and background on Massive MIMO channels.
\begin{definition}[Gaussian Random Matrix]
A real or complex Gaussian $n\times n$ matrix $\mathbf{H}$
has independent and identically distributed (i.i.d.) real or complex zero-mean Gaussian entries with identical variance $1/n$.
\end{definition}

\begin{definition}[SVD and Truncated SVD]
Let \( \mathbf{H} \in \mathbb{C}^{m \times n} \) be a matrix of rank \( r \), with singular value decomposition (SVD)
\[
\mathbf{H} = \mathbf{U} \boldsymbol{\Sigma} \mathbf{V}^H,
\]
where \( \mathbf{U} \in \mathbb{C}^{m \times m} \) and \( \mathbf{V} \in \mathbb{C}^{n \times n} \) are unitary matrices, $\mathbf{V}^H$ is the conjugate transpose of $\mathbf{V}$, and \( \boldsymbol{\Sigma} \in \mathbb{R}^{m \times n} \) is a diagonal matrix whose diagonal entries \( \sigma_1 \geq \sigma_2 \geq \cdots \geq \sigma_r > 0 \) are the singular values of \( \mathbf{H} \). The \emph{truncated SVD} of rank \( k \leq r \) is
\[
\mathbf{H}_k = \sum_{i=1}^{k} \sigma_i \mathbf{u}_i \mathbf{v}_i^H = \mathbf{U}_k \boldsymbol{\Sigma}_k \mathbf{V}_k^H,
\]
where \( \mathbf{U}_k \in \mathbb{C}^{m \times k} \) contains the first \( k \) columns of \( \mathbf{U} \),  \( \boldsymbol{\Sigma}_k \in \mathbb{R}^{k \times k} \) is a diagonal matrix with the top \( k \) singular values, \( \mathbf{V}_k \in \mathbb{C}^{n \times k} \) contains the first \( k \) columns of \( \mathbf{V} \).
\end{definition}

\subsection{MIMO Channel Model} 
We consider an $n \times n$ MIMO channel:
\begin{equation}
\setlength{\abovedisplayskip}{3pt}
\setlength{\belowdisplayskip}{3pt}
    \mathbf{y}=\sqrt{n}\mathbf{H} \cdot \mathsf{Enc}(\mathbf{s})+\mathbf{z}=\sqrt{n}\mathbf{H} \mathbf{x}+\mathbf{z}, \label{MIMO_Ch}
\end{equation}
where $\mathbf{H}$ is an $n \times n$ complex Gaussian random matrix, $\mathbf{s}=[s_1,\ldots,s_n]^T$ contains the information symbols, the function $\mathsf{Enc}(\mathbf{s})$ encodes $\mathbf{s}$ into a transmitted vector $\mathbf{x}=[x_1,\ldots,x_n]^T$, and $\mathbf{y}=[y_1,\ldots,y_n]^T$ is the received vector. The components of $\mathbf{z} = [z_1, \ldots, z_n]^T$, are i.i.d. complex-valued Gaussian noise with zero mean and variance of $\sigma^2$. 
 We assume that the complex-valued symbols $s_i=\mathfrak{R}(s_i)+\mathfrak{I}(s_i) \sqrt{-1}$ satisfy
\begin{equation}
\setlength{\abovedisplayskip}{3pt}
\setlength{\belowdisplayskip}{3pt}
    \mathbb{E}(\mathfrak{R}(s_i)^2)= \mathbb{E}(\mathfrak{I}(s_i)^2)=1, \; \text{for} \; 1 \leq i \leq n. \label{n_signal}
\end{equation}
We also impose an average power constraint $\mathbb{E}(\|\mathbf{x}\|^2) = P$. The signal-to-noise ratio (SNR) is denoted as $P/\sigma^2$.

\subsection{SVD Precoding and Power Allocation Schemes} \label{Sec:II-B}
We recall the SVD-based precoding scheme in \cite{Tse05}. Considering the SVD decomposition of $\mathbf{H}=\mathbf{U}\mathbf{\Sigma}\mathbf{V}^H$, We choose the encoding function in \eqref{MIMO_Ch} as 
\begin{equation}
 \mathbf{x}=\mathsf{Enc}(\mathbf{s})= \mathbf{V}\cdot (\mathbf{p} \odot\mathbf{s}) =  \mathbf{V}\cdot \hat{\mathbf{s}}, \label{CP_ENC}   
\end{equation}
where $\odot$ represents the element-wise product, $\mathbf{p}=[\sqrt{p_1/2}, \ldots, \sqrt{p_n/2}]^T$ is the power allocation vector whose entries satisfy $\sum_{i=1}^n p_i=P$, and  $ \hat{\mathbf{s}} = \mathbf{p} \odot\mathbf{s} =[\hat{s}_1,\ldots,\hat{s}_n]^T$. With full CSI, the received vector can be rewritten as
\begin{eqnarray}
   \mathbf{y}' =\mathbf{U}^{-1}\mathbf{y}=\sqrt{n}\mathbf{\Sigma}\cdot \hat{\mathbf{s}}+\mathbf{z}', \label{MIMO_SIC}
\end{eqnarray}
where $\mathbf{y}'= [y'_1,\ldots,y'_n]^T$ and $\mathbf{z}'=\mathbf{U}^{-1}\mathbf{z}= [z'_1,\ldots,z'_n]^T$. The entries of $\mathbf{z}'$ are i.i.d. complex-valued Gaussian noise with zero mean and variance of $\sigma^2$. In summary, the SVD decomposition converts the channel \eqref{MIMO_Ch} into $n$ parallel subchannels:
\begin{equation}
 {y}'_{i}=\sqrt{n}R_{i,i}\cdot \hat{s}_{i}+z'_{i}, \; \text{for} \; 1\leq i \leq n, \label{Pal_CH}   
\end{equation}
where ${R}_{i,i}$ is the $i^{\text{th}}$ diagonal element in $\mathbf{\Sigma}$.

With Gaussian inputs, i.e., $s_i \in \mathcal{CN}(0,1)$, the capacity of \eqref{MIMO_SIC} is achieved by the \emph{waterfilling}  (WF) power allocation \cite{Tse05}:
\begin{equation}
C_{\mathsf{Gaussian}} =\sum_{i=1}^{n}\log\left( 1 + \dfrac{p_i}{\eta_i}\right), \; \text{where}\label{capacity_gauss}
\end{equation}
\begin{equation}
    p_i =(\lambda_{\mathsf{WF}}-\eta_i)_{+}, \; \text{for} \; 1 \leq i \leq n, \label{WF_P}
\end{equation}
$(x)_{+}=\max\{0,x\}$ and $\eta_{i}=\sigma^2/(n{R}_{i,i}^2)$ is the normalized noise power for the $i^{\text{th}}$ subchannel in \eqref{MIMO_SIC}. All logarithms are in base $2$. The Lagrange multiplier $\lambda_{\mathsf{WF}} > 0$ is found as a unique solution for $\sum_{i=1}^{n}p_i(\lambda_{\mathsf{WF}})=P$.

In this paper, we consider employing a uniform \( M_i \)-QAM constellation for each \( s_i \), where \( M_i \) is a power of $4$. Under the condition in \eqref{n_signal}, the $M_i$-QAM constellation is given by
\begin{equation*}
\setlength{\abovedisplayskip}{3pt}
\setlength{\belowdisplayskip}{3pt}
\{\mathfrak{R}(s_i), \mathfrak{I}(s_i)\} \subseteq \left\{(2\ell-\sqrt{M_i}-1)\sqrt{\dfrac{3}{M_i-1}}, 1\leq \ell \leq \sqrt{M_i}\right\}.
\end{equation*}

With  \( \{M_i\}_{i=1}^{n} \)-QAM constellations, the capacity of \eqref{MIMO_SIC} is achieved by the \emph{Mercury/Waterfilling} (MWF) \cite{MWF2006}. We recall the approximate closed-form solution for MWF in \cite{MWF_App2021}:
\begin{equation}
C_{\mathsf{QAM}} \approx \sum_{i=1}^{n}\left(\log\left( 1 + \dfrac{p_i}{\eta_i}\right) -\log\left( 1 + \dfrac{p_i}{\eta_i M_i}\right) \right),  \label{capacity_QAM}
\end{equation}
where 
\begin{equation}
 p_i = \dfrac{\eta_i}{2} \left ( \sqrt{(M_i-1)^2+\dfrac{4}{\eta_i\lambda_{\mathsf{MWF}}}(M_i-1)}-M_i-1\right)_{+}, \label{mw_pk}
\end{equation}
for $1 \leq i \leq n$, and the Lagrange multiplier $\lambda_{\mathsf{MWF}} > 0$ is found as a unique solution for $\sum_{i=1}^{n}p_i(\lambda_{\mathsf{MWF}})=P$.

With \( \{M_i\}_{i=1}^{n} \)-QAM constellations, the \emph{error/waterfilling} (EWF) strategy from \cite{EWF2003} offers an alternative power allocation aimed at minimizing total decoding error. At high \(\mathsf{SNR}\) with Gray mapping, a symbol error typically alters only one bit, allowing the bit error rate (BER) to be approximated as the symbol error rate (SER) divided by the bits per symbol. Thus, the BER for the \(i^{\text{th}}\) subchannel in \eqref{Pal_CH} is given by:
\begin{equation}
\mathsf{BER}_i=  \dfrac{4}{\log M_i}\left(1-\dfrac{1}{\sqrt{M_i}} \right)Q\left(\sqrt{\dfrac{p_i}{\eta_i}\cdot \dfrac{3}{M_i-1}}\right). \label{BER_QAM}
\end{equation}
Applying the union bound, we define the EWF problem below.
\begin{definition}[EWF Power Allocation Problem]~\label{Def.EWFP} The EWF problem is defined by
\begin{equation}
\{p_i\}_{i=1}^n = \arg \min_{p_1'+\cdots+p_{n}'= P}\sum_{i=1}^{n}\mathsf{BER}_i(M_i,p_i'),  \label{BER_EWF}
\end{equation}
where $\mathsf{BER}_i$ is given in \eqref{BER_QAM}.
\end{definition}

It is important to note that the above EWF formulation differs from the original EWF in \cite{EWF2003}, which aims to minimize the total SER. In contrast, our approach targets the reduction of the total BER, a metric that directly influences the accuracy of data transmission at the bit level. For completeness, the solution to the EWF problem of Def.~\ref{Def.EWFP} is provided below.

\begin{theo}\label{Theo:EMF_Solution}
The optimal power allocation for the EWF problem is unique and satisfy
\begin{equation}
p_i=\dfrac{W((A_i\cdot \lambda_{\mathsf{EWF}})^{-2})}{B_i} , \; \text{for} \; 1\leq i \leq n  \label{EWF_p}
\end{equation}
where $W(x)$ is the real-valued Lambert W function defined
as the inverse of the function $f(w)=w\exp(w)$ for $w>0$, and
\begin{equation}
    B_i= \dfrac{3}{(M_i-1)\eta_i}, \; A_i=\dfrac{\sqrt{2\pi M_i}\log(M_i)}{2B_i\cdot (\sqrt{ M_i}-1)}. 
\end{equation}
The Lagrange multiplier $\lambda_{\mathsf{EWF}} > 0$ is found as a unique solution for $\sum_{i=1}^{n}p_i(\lambda_{\mathsf{EWF}})=P$.
\end{theo}

\begin{IEEEproof}
The proof follows the Lagrange multiplier method and is similar to that in \cite[Sec. III-A]{EWF2003}, so it is omitted.
\end{IEEEproof}

\subsection{Adaptive QAM (A-QAM)}
Although power allocation and MIMO techniques have been extensively studied in the literature \cite{LaiLiang2015}\cite{Jinhong2019}\cite{Jingge2024}\cite{Song2020}\cite{PA_survey2018}\cite{MMcKay2015}\cite{LatticePA2024}, the optimal joint design involving power allocation, constellation size selection, and bit-allocation remains an open problem. As demonstrated in Section \ref{Sec:II-B}, the choice of the QAM sizes directly influences the power distribution, thereby affecting the achievable QAM capacity, \( C_{\mathsf{QAM}} \). Several approaches for adaptive QAM (A-QAM) size selection have been proposed. For example, we recall the method in \cite{Palomar2005}:
\begin{equation}
    M_i \approx 1 + \dfrac{p_i}{\eta_i \Gamma}, \; \text{for} \; 1\leq i \leq n, \label{QAM_peWF}
\end{equation}
where $\Gamma=(2/3)\log_{e}(2/\mathsf{SER}_i)$, and $\mathsf{SER}_i$ is a predefined $\mathsf{SER}$ for the $i^{\text{th}}$ subchannel in \eqref{Pal_CH}.
\cite{Palomar2005} proposes an power allocation strategy aimed at maximizing the total transmitted bit count, as opposed to maximizing mutual information, i.e.,
\begin{equation}
\{p_i\}_{i=1}^n = \arg \max_{p_1'+\cdots+p_{n}'= P}\sum_{i=1}^{n}\log M_i(p_i'), 
\end{equation}
whose solution also follows a WF solution
\begin{equation}
 p_i =(\lambda_{\mathsf{SER\_WF}}-\eta_i\Gamma)_{+}, \; \text{for} \; 1 \leq i \leq n. \label{Pe_WF}
\end{equation}
The Lagrange multiplier $\lambda_{\mathsf{SER\_WF}} > 0$ is found as a unique solution for $\sum_{i=1}^{n}p_i(\lambda_{\mathsf{SER\_WF}})=P$.

By comparing \eqref{Pe_WF} and \eqref{WF_P}, it is observed that for uniformly low \(\mathsf{SER}_i\) (e.g., \(10^{-3}\)), the method in \cite{Palomar2005} tends to deactivate significantly more subchannels than the Gaussian-input case, leading to a substantial rate loss relative to $C_{\mathsf{Gaussian}}$.

\subsection{Problem Statement}

Power allocation strategies often recommend deactivating a subset of subchannels with poor channel conditions. The number of such inactive subchannels is defined as
\begin{equation}
 k_{\text{opt}}:=  \sum_{i=1}^{n} {1}_{\{p_i = 0\}},
\end{equation}
where \(\{p_i = 0\}_{i=1}^n\) characterizes a non-trivial power allocation scheme, such as WF, MWF, or EWF. If \( k_{\text{opt}} > 0 \) almost surely, then the design problem reduces to determining \( \{M_i\}_{i=1}^{n-k_{\text{opt}}} \). Moreover, since \( k_{\text{opt}} \) is a function of the singular values of \( \mathbf{H} \), it can be approximated by a constant for a large-scale \( \mathbf{H} \) \cite[Theorem 2.37]{Verdu04}. In such case, truncated SVD can replace the full SVD with reduced complexity \( O(n(n - k_{\text{opt}})^2) \).

\begin{definition}[Non-Zero Subchannel Deactivation Problem]\label{Def.kopt} For a given $\mathsf{SNR}$ and power allocation scheme, determine if 
\begin{equation}
  \lim_{n\rightarrow \infty} k_{\text{opt}} = \lim_{n\rightarrow \infty} \sum_{i=1}^{n} {1}_{\{p_i = 0\}}>0.
\end{equation}     
\end{definition}

We further aim to jointly design A-QAM and power allocation, avoiding the rate loss due to a very large $k_{\text{opt}}$.
\begin{definition}[Joint QAM Size Selection and Power Allocation Problem]\label{Def.JPB}
Given a target $\mathsf{SNR}$, the joint optimization problem seeks to determine the constellation sizes \( \{M_i\}_{i=1}^{n} \) and corresponding power distribution \( \{p_i\}_{i=1}^{n} \) such that the resulting achievable capacity \( C_{\mathsf{QAM}} \) satisfies the constraint:  
\begin{equation}
    C_{\mathsf{Gaussian}} - C_{\mathsf{QAM}} \leq n.
\end{equation}
Note that $C_{\mathsf{Gaussian}}$ in \eqref{capacity_gauss} remains constant for a given $\mathsf{SNR}$.
\end{definition}

\section{Non-Zero Subchannel Deactivation}
This section addresses the problem in Def.~\ref{Def.kopt}, showing that in massive MIMO, truncated SVD precoding matches full SVD performance with substantially lower complexity.

\begin{lemma}[$k_{\text{opt}}$ with WF]\label{lem:k_WF} With a fixed $\mathbf{SNR}$, it holds
\begin{equation}
  \lim_{n\rightarrow \infty} k_{\text{opt}} >0, \; \text{for the WF}.
\end{equation}
\end{lemma}

\begin{IEEEproof} We assume $k_{\text{opt}}=0$ and prove the result by contradiction. If $k_{\text{opt}}=0$, it follows from \eqref{WF_P} that
 \begin{equation}
 p_i=\lambda_{\mathsf{WF}}-\eta_i >0, \; \text{for} \; 1 \leq i \leq n. 
 \end{equation}
For the weakest subchannel with index $i=n$, it holds $\lambda_{\mathsf{WF}} =\infty$, since \(\lim_{n\rightarrow \infty} \eta_n=\infty\) \cite[Theorem 1.1]{Tao2010}. Recalling that $\lim_{n \rightarrow \infty} \eta_1=0$ \cite[Theorem 2.37]{Verdu04}, we have
\begin{equation}
   \lim_{n\rightarrow \infty} p_1= \lambda_{\mathsf{WF}}= \infty.
\end{equation}
This contradicts the condition $p_1\leq P$.
\end{IEEEproof}
It is worth noting that Lemma \ref{lem:k_WF} remains applicable to the method in \cite{Palomar2005}, as \(\Gamma > 1\) in \eqref{Pe_WF} holds whenever \(\mathsf{SER}_i \leq 10^{-1}\).
 
\begin{lemma}[$k_{\text{opt}}$ with MWF]\label{lem:k_MWF} With a fixed $\mathbf{SNR}$, it holds
\begin{equation}
  \lim_{n\rightarrow \infty} k_{\text{opt}} >0, \; \text{for the MWF}.
\end{equation}
\end{lemma}

\begin{IEEEproof} We assume $k_{\text{opt}}=0$ and prove the result by contradiction. If $k_{\text{opt}}=0$, it holds \cite[Theorem 1]{MWF_App2021}
 \begin{equation}
   \lambda_{\mathsf{MWF}} = \dfrac{M_i-1}{p_i^2/\eta_i+(M_i+1)p_i+  M_i\eta_i},  \; \text{for} \; 1\leq i \leq n.
 \end{equation}
For the weakest subchannel with index $i=n$, it holds $\lambda_{\mathsf{MWF}} =0$, since \(\lim_{n\rightarrow \infty} \eta_n=\infty\) \cite[Theorem 1.1]{Tao2010}. This contradicts the condition \(\lambda_{\mathsf{MWF}} > 0\) in \cite[Theorem 1]{MWF_App2021}.
\end{IEEEproof}

\begin{lemma}[$k_{\text{opt}}$ with EWF]\label{lem:k_EWF} With a fixed $\mathbf{SNR}$, it holds
\begin{equation}
  \lim_{n\rightarrow \infty} k_{\text{opt}} >0,  \; \text{for the EWF}.
\end{equation}
\end{lemma}

\begin{IEEEproof} We assume \(\lim_{n\rightarrow \infty} k_{\text{opt}}=0\) and prove the result by contradiction. If \(\lim_{n\rightarrow \infty}k_{\text{opt}}=0\), then for all \( 1 \leq i \leq n \), we have \( p_i > 0\). Focusing on the weakest subchannel indexed by \( i = n \), and using \eqref{EWF_p}, we obtain:
\begin{equation}
\lambda_{\mathsf{EWF}}= \dfrac{4}{\log M_n}\left(1-\dfrac{1}{\sqrt{M_n}} \right) \dfrac{\exp(-p_nB_n/2)}{\sqrt{2\pi}} \dfrac{\sqrt{B_n}}{2\sqrt{p_n}}.
\end{equation}
Since \(\lim_{n\rightarrow \infty} B_n=0\) \cite[Theorem 1.1]{Tao2010}, and given that \( 0< p_n \leq P \), it follows that \(\lim_{n\rightarrow \infty} \lambda_{\mathsf{EWF}} =0\). This contradicts the condition \(\lambda_{\mathsf{EWF}} > 0\) in Theorem \ref{Theo:EMF_Solution}.
\end{IEEEproof}

Lemmas \ref{lem:k_WF}, \ref{lem:k_MWF}, and \ref{lem:k_EWF} directly yield the following theorem.
\begin{theo}
For fixed \(\mathsf{SNR}\) and large \(n\), SVD precoding under WF, MWF, or EWF reduces to transmission over the \(n - k_{\text{opt}}\) strongest subchannels, i.e., truncated SVD precoding:
\begin{equation}
 {y}'_{i}=\sqrt{n}R_{i,i}\cdot \hat{s}_{i}+z'_{i}, \; \text{for} \; 1\leq i \leq n-k_{\text{opt}}, \label{Pal_CH_K}   
\end{equation}
where $k_{\text{opt}}>0$ and can be calculated from \eqref{WF_P}, \eqref{mw_pk}, or \eqref{EWF_p}.
\end{theo}

\begin{remark}
The value of \( k_{\text{opt}} \) can be relatively large for a finite $n$, e.g., \( k_{\text{opt}} = 13 \) in a \( 32 \times 32 \) MIMO system employing the A-QAM scheme \cite{Palomar2005}. Further details are provided in Section \ref{sec:IV-A}. While a large \( k_{\text{opt}} \) reduces the total precoding complexity via truncated SVD, it may lead to a loss in data rate due to the deactivation of a substantial number of subchannels. In the following section, we propose an A-QAM scheme that preserves a constant gap to \( C_{\text{Gaussian}} \) in \eqref{capacity_gauss}, for a given $n$.  
\end{remark}

\section{Adaptive QAM and Bit-Allocation}
In this section, we will present a joint power and bit allocation design addressing Def.~\ref{Def.JPB}. 
\subsection{The Proposed Adaptive QAM} \label{sec:IV-A}
\begin{lemma}[From MWF to WF]\label{lem: M_i} Given the optimal $\{p_i\}_{i=1}^n$ obtained from the traditional WF, if $M_i \approx p_i/\eta_i$, then 
\begin{equation}
C_{\mathsf{QAM}} \approx \sum_{i=1}^{n} \left(\log\left( 1 + \dfrac{p_i}{\eta_i}\right) -1\right)_{+}, \label{cap_app}
\end{equation}
i.e., the capacity with discrete input constellations is achieved using the conventional WF strategy and A-QAM.
\end{lemma}

\begin{IEEEproof}
With $M_i \approx p_i/\eta_i$, \( C_{QAM} \) in \eqref{capacity_QAM} reduces to \eqref{cap_app}, since the power distribution \( \{p_i\}_{i=1}^n \) from the traditional WF also maximizes the sum rate in the high-\(\mathsf{SNR}\) regime.
\end{IEEEproof}

\begin{remark}
Lemma \ref{lem: M_i} establishes a reduction from the MWF to the classical WF through the adaptive choice of \(\{M_i \approx p_i/\eta_i\}_{i=1}^n\), resulting in an approximate capacity loss characterized by \( C_{\mathsf{Gaussian}} - C_{\mathsf{QAM}} \approx n - k_{\text{opt}} \), where \( k_{\text{opt}} \) denotes the number of subchannels deactivated under the WF solution. Furthermore, this reduction simplifies the computation of the MWF power distribution in \eqref{mw_pk}, eliminating the need for iterative numerical methods such as bisection search in \cite{MWF_App2021}.
\end{remark}

\begin{figure}[tbp]
\centering
\includegraphics[width=0.47\textwidth]{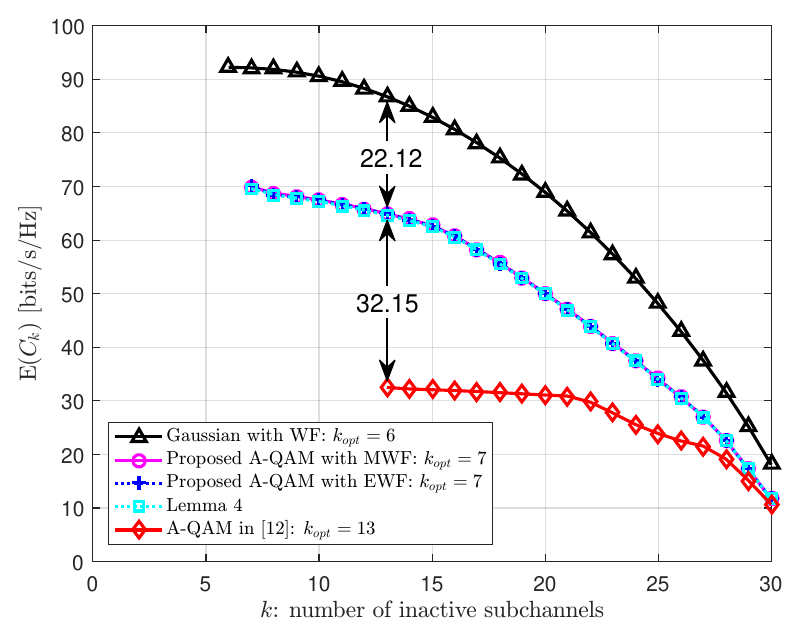} \vspace{-3mm} 
\caption{\( 32 \times 32 \) MIMO at \( \mathsf{SNR} = 10 \) dB: Average capacity \( \mathbb{E}(C_k) \) for different input types and power allocation schemes, with \( k \) inactive weakest subchannels.}
\label{fig:com_capacity_MEWF}
\vspace{-3mm}
\end{figure}

Fig.~\ref{fig:com_capacity_MEWF} shows the average capacity \( \mathbb{E}(C_k) \) of a \( 32 \times 32 \) MIMO system versus the number of deactivated subchannels \( k \), at \( \mathsf{SNR} = 10 \) dB (\( \sigma^2 = 6.4, P = 64 \)). For Gaussian inputs, conventional WF is used to maximize mutual information. For QAM inputs, the proposed A-QAM scheme is evaluated, where constellation sizes \( \{M_i\}_{i=1}^n \) are selected using Lemma \ref{lem: M_i}, followed by either MWF (for mutual information maximization) or EWF (for BER minimization). Capacity is maximized by deactivating \( k_{\text{opt}} > 0 \) subchannels, as predicted by Theorem \ref{Theo:EMF_Solution}. The capacity estimate from Lemma \ref{lem: M_i}, computed via \eqref{capacity_QAM} using WF power distribution in \eqref{WF_P}, closely matches the MWF curve, validating the reduction for MWF to WF under the proposed A-QAM scheme. The EWF curve also aligns closely, confirming that BER improvements at fixed \( \mathsf{SNR} \) do not incur capacity loss under the proposed A-QAM scheme.

For comparison, we evaluate the maximum achievable rate of the A-QAM scheme from \cite{Palomar2005}, fixing \(\mathsf{SER}_i = 10^{-3}\) and computing the rate via \eqref{capacity_QAM} using the power allocation \(\{p_i\}_{i=1}^n\) from \eqref{Pe_WF} and corresponding modulation orders \(\{M_i\}_{i=1}^n\) from \eqref{QAM_peWF}. The simulation result indicates that the A-QAM in \cite{Palomar2005} achieves a notably lower rate than the proposed method.

These results emphasize that different power allocation strategies and QAM constellation selections produce varying \( k_{\text{opt}} \) values, exhibiting distinct complexity–rate tradeoffs.

\subsection{Bit-Allocation for a Specified Transmission Data Rate $R$}\label{Sec:IV-B}
The MWF is closely related to MMSE decoding \cite{MWF2006}. At high \(\mathsf{SNR}\), the MWF strategy balances the mean squared error (MSE) across subchannels, reducing the overall decoding error. Together with Fig.~\ref{fig:com_capacity_MEWF}, which shows that EWF with the proposed A-QAM scheme achieves no rate loss, this motivates a bit-allocation design that explicitly minimizes total decoding error, favoring reliability over mutual information optimality. 

Given an \(\mathsf{SNR}\), let \(\hat{R}\ = \sum_{i=1}^n
\log M_i \) be the total number of transmitted
bits. As \(\hat{R}\) may differ from the target transmission data rate \(R\), we iteratively adjust \(\{M_i\}_{i=1}^{n}\) based on \(\{\mathsf{BER}_i\}_{i=1}^{n}\) from \eqref{BER_QAM}. At each iteration, if \(\hat{R} < R\), we increase the QAM size of the best subchannel; if \(\hat{R} > R\), we reduce that of the worst. The pseudocode is given in Algorithm~\ref{alg:QAM_Update}.

\begin{lemma}[Worst-Case BER Minimization]\label{lem:bitAlo} Consider an alternative iterative bit-allocation algorithm $\mathcal{B}$ that employs a different $1$-bit allocation criterion from that of Algorithm~\ref{alg:QAM_Update} in each iteration. Let $\{\hat{M}_i\}_{i=1}^n$ denote the output of $\mathcal{B}$, and let $\{M_i\}_{i=1}^n$ represent the output of Algorithm~\ref{alg:QAM_Update}. It holds
\begin{equation}
    \max\{\{\text{BER}_i(M_i)\}_{i=1}^n\} \leq \max\{\{\text{BER}_i(\hat{M}_i)\}_{i=1}^n\}.
\end{equation} 
\end{lemma}
\begin{IEEEproof}
We begin by considering the case where \(\hat{R} > R\). Without loss of generality, we assume that from the \(j^{\text{th}}\) iteration onward, the sets \(\{M_i^{(j)}\}_{i=1}^n \neq \{\hat{M}_i^{(j)}\}_{i=1}^n\), where \(j\) denotes the output at the \(j^{\text{th}}\) iteration. By referring to Lines 8–10 of Algorithm \ref{alg:QAM_Update}, the following inequality holds:
\begin{equation}
\max\{\{ \text{BER}_i(M_i^{(j)}) \}_{i=1}^n\} \leq \max\{\{\text{BER}_i(\hat{M}_i^{(j)})\}_{i=1}^n \}. \label{itr_j}
\end{equation}
For the \((j+1)^{\text{th}}\) iteration, from \eqref{itr_j}  and Lines $8-10$, it holds:
\begin{equation*}
\max\{\{ \text{BER}_i(M_i^{(j+1)}) \}_{i=1}^n\} \leq \max\{\{\text{BER}_i(\hat{M}_i^{(j+1)})\}_{i=1}^n \}.
\end{equation*}
By induction, this property holds for all $|b|\geq j$. A similar argument can be applied to the case of \(\hat{R} < R\).
\end{IEEEproof}

\vspace{-2mm}

\begin{algorithm}[H]
\caption{$\mathsf{BitAllocation}(\{M_i\}_{i=1}^{n},\{\mathsf{BER}_i\}_{i=1}^{n},R)$}
\label{alg:QAM_Update}
\begin{algorithmic}[1]

	\State
	$b = \sum_{i=1}^{i=n}\log M_i -R$
	
\For{$j= 1 : 1 : |b| $}
 
\If{$b < 0 $}   \hfill{//increase QAM size}

        \State $t =arg\min_{1\leq i \leq n}\{\{\mathsf{BER}_i({M_i})\}_{i=1}^{n}\}$      \hfill{//BST}

        \State $M_t =4M_t$
        \State $\mathsf{BER}_t =\mathsf{BER}_t({M_t})$

 \ElsIf {$b > 0$}         \hfill{//decrease QAM size}

        \State $t =arg\max_{1\leq i \leq n}\{\{\mathsf{BER}_i({M_i})\}_{i=1}^{n}\}$      \hfill{//BST}

        \State $M_t =M_t/4$
        \State $\mathsf{BER}_t =\mathsf{BER}_t({M_t})$

        \If{$M_t == 1 $}               \hfill{//deactivate the $t^{\text{th}}$ subchannel}
         \State $\mathsf{BER}_t =0$

         \EndIf

 \EndIf
\EndFor
	
	\State \Return $\{M_i\}_{i=1}^{n}$

\end{algorithmic}
\end{algorithm}

\vspace{-2mm}

\begin{remark}
 Algorithm \ref{alg:QAM_Update} can be applied to any predefined set of constellations \(\{M_i\}_{i=1}^{n}\). If, for some \(1 \leq t \leq n\), we have \(M_t = 1\), this indicates that the \(t^{\text{th}}\) subchannel is deactivated. In addition to determining \(k_{\text{opt}}\) using either MWF or EWF, Algorithm \ref{alg:QAM_Update} also identifies the indices of additional subchannels that must be deactivated due to the fixed rate, i.e.,
 \begin{equation}
 k_{\text{opt}}=     \sum_{i=1}^{n} {1}_{\{p_i = 0 \; \text{or} \; M_i=1\}}. 
 \end{equation}
The complexity of Algorithm \ref{alg:QAM_Update} is \(O(|R - \hat{R}| \log n)\), utilizing a balanced binary search tree (BST) to perform the minimum/maximum operation during each iteration. When the initial input \(\{M_i\}_{i=1}^{n}\) is based on Lemma \ref{lem: M_i}, we have \(\hat{R} \approx \sum_{i=1}^n \log(p_i/\eta_i)\). If \(R\) is selected close to the corresponding \(C_{\text{QAM}}\) in \eqref{cap_app}, the complexity is bounded by \(O(n \log n)\) in the high-\(\mathsf{SNR}\) regime. Lemmas \ref{lem: M_i} and \ref{lem:bitAlo} directly yield the following theorem addressing the problem set in Def \ref{Def.JPB}.
 \end{remark}

 \begin{theo} For a given $\mathsf{SNR}$, the proposed A-QAM scheme yields an approximate capacity gap of \( C_{\mathsf{Gaussian}} - C_{\mathsf{QAM}} \approx n - k_{\text{opt}} \), where \( k_{\text{opt}} \) is obtained from \eqref{WF_P}. Combined with the proposed bit-allocation algorithm, the worst-case BER is minimized with complexity \(O(n\log n) \).
 \end{theo}

\begin{figure}[tbp]
\centering
\includegraphics[width=0.47\textwidth]{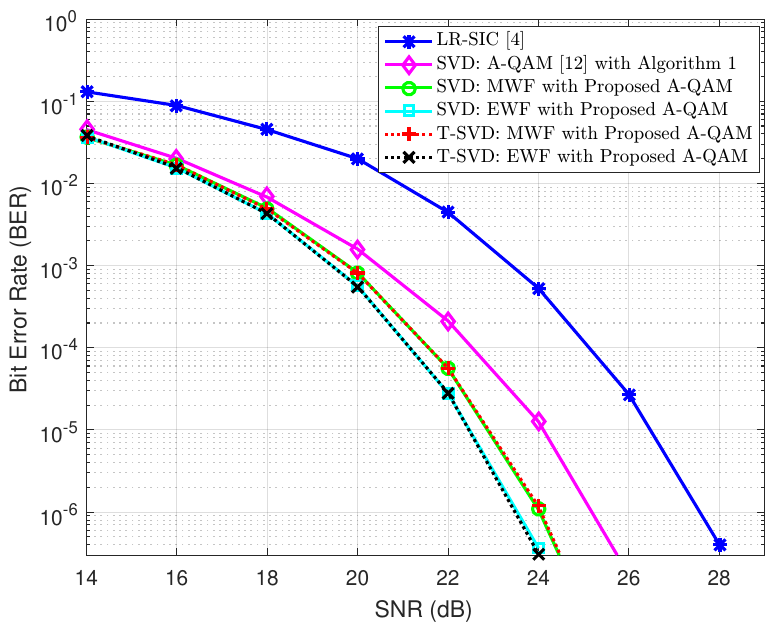} \vspace{-3mm} 
\caption{$96 \times 96$ MIMO: BER of different precoding schemes.}
\label{fig:com_ber_EWF96}
\vspace{-5mm}
\end{figure}

Fig. \ref{fig:com_ber_EWF96} compares the BER performance of full SVD precoding and the proposed truncated SVD precoding (T-SVD), using \( \{M_i\}_{i=1}^{n - k_{\text{opt}}} \) derived from Lemma \ref{lem: M_i} and Algorithm \ref{alg:QAM_Update}, under both MWF and EWF frameworks. A \(96 \times 96\) MIMO system is considered, with total transmit power \(P = 192\) and a target data rate of \(R = 384\) bits. In the T-SVD precoding, for each value of \( \mathsf{SNR} \),  \( \{M_i\}_{i=1}^{n - k_{\text{opt}}} \) are precomputed based on the singular value distribution in \cite[Theorem 2.37]{Verdu04}. We notice that the value of $k_{\text{opt}}$ is non-negligible, for instance, under EWF, $k_{\text{opt}}=14$ at \( \mathsf{SNR} =22 \) dB. T-SVD is then employed to create the \( n - k_{\text{opt}} \) strongest subchannels. In contrast, full SVD precoding dynamically determines \( \{M_i\}_{i=1}^{n-k_{\text{opt}}} \) based on the instantaneous channel realization. Despite this difference, simulation results indicate that T-SVD precoding achieves BER performance nearly identical to that of full SVD precoding. For reference, we include LR-SIC precoding from \cite{Windpassinger04} with 16-QAM across all $96$ subchannels, and the A-QAM scheme from \cite{Palomar2005}, using a uniform target \(\mathsf{SER}_i = 10^{-3}\) and Algorithm \ref{alg:QAM_Update}. Results show T-SVD precoding with EWF consistently outperforms other schemes. Notably, the respective computational complexities of T-SVD, SVD, and LLL reduction in \cite{Windpassinger04} are \( O(n(n - k_{\text{opt}})^2) \), $O(n^3)$, and \( O(n^3\log n) \) \cite{LLLComplexityMIMO2013}. These results affirm that the proposed joint power and bit allocation effectively reduces both BER and precoding complexity.

\section{Conclusion}
We have demonstrated that, in massive MIMO channels, a non-trivial power allocation strategy necessarily deactivates a subset of weak subchannels to enhance system performance, enabling the use of truncated SVD in place of full SVD precoding. Moreover, integrating truncated SVD with the proposed EWF framework, adaptive QAM, and bit allocation leads to improved decoding performance and reduced precoding complexity compared to conventional approaches. Future work will extend this framework to cases with imperfect CSIT.

\bibliographystyle{IEEEtran}
\bibliography{IEEEabrv,LIUBIB}

\end{document}